\documentclass[prb,onecolumn, amsmath,amssymb,preprintnumbers,showpacs,nofootinbib]{revtex4}
\usepackage{latexsym}
\usepackage{epsfig}
\usepackage{amssymb}
\usepackage{amsmath}
\usepackage{color}

\newcommand{\bea}{\begin{eqnarray}}
\newcommand{\eea}{\end{eqnarray}}
\newcommand{\la}{\langle}
\newcommand{\ra}{\rangle}

\newcommand{\potent}{\Phi_{\rm potent}}
\newcommand{\impotent}{\Phi_{\rm impotent}}
\newcommand{\badactors}{\Phi_{\rm badactors}}

\newcommand{\average}[1]{\langle #1 \rangle}

\begin{document}

\title{Teaching the Principles of Statistical Dynamics}
\author{Kingshuk Ghosh}
\author{Ken A. Dill}
\email{dill@maxwell.ucsf.edu}
\affiliation{Department of Biophysics, University of California, San Francisco
CA 94143}
\author{Mandar M. Inamdar}
\author{Effrosyni Seitaridou}
\author{Rob Phillips$^{\rm 1}$}
\email{phillips@pboc.caltech.edu}
\affiliation{Division of Engineering and Applied Science and
$^{\rm 1}$ Kavli Nanoscience Institute, California Institute of Technology,
Pasadena, CA 91125}

\begin{abstract}
We describe a simple framework for teaching the principles that
underlie the dynamical laws of transport: Fick's law of diffusion,
Fourier's law of heat flow, the Newtonian viscosity law, and
mass-action laws of chemical kinetics. In analogy with the way that
the maximization of entropy over microstates leads to the Boltzmann
law and predictions about equilibria, maximizing a quantity that E. T.
Jaynes called ``Caliber" over all the possible {\it microtrajectories}
leads to these dynamical laws. The principle of Maximum Caliber also
leads to dynamical distribution functions which characterize the
relative probabilities of different microtrajectories. A great source
of recent interest in statistical dynamics has resulted from a new
generation of single-particle and single-molecule experiments which
make it possible to observe dynamics one trajectory at a time.
\end{abstract}


\maketitle

\section{Introduction}
We describe an approach for teaching the principles that underlie the
dynamical laws of transport: of particles (Fick's law of diffusion),
energy (Fourier's law of heat flow), momentum (the Newtonian law for
viscosity),~\cite{rief65} and mass-action laws of chemical
kinetics.~\cite{dill03} Recent experimental advances now allow for
studies of forces and flows at the single-molecule and nanoscale
level, representative examples of which may be found in the
references.~\cite{kasian96, lu98, rief00, meller01,smith01, li02,
liphardt02, bustamante03} For example, single-molecule methods have
explored the packing of DNA inside viruses,~\cite{smith01} and the
stretching of DNA and RNA molecules.~\cite{liphardt02,bustamante03}
Similarly, video microscopy now allows for the analysis of
trajectories of individual submicron size colloidal
particles~\cite{dufresne01}, and the measurement of single-channel
currents has enabled the kinetic studies of DNA translocation through
nanopores.~\cite{kasian96, meller01}

One of the next frontiers in biology is to understand the ``small
numbers'' problem: how does a biological cell function, given that
most of its proteins and nucleotide polymers are present in numbers
much smaller than Avogadro's number?~\cite{alberts97} For example, one
of the most important molecules, a cell's DNA, occurs in only a single
copy. Also, it is the flow of matter and energy through cells that
makes it possible for organisms to maintain a relatively stable
form.~\cite{kondepudi98} Hence, in order to function, cells always
have to be in this state far from equilibrium. Thus, many problems of
current interest involve small systems that are out of equilibrium.
Our interest here is two-fold: to teach our students a physical
foundation for the phenomenological macroscopic laws, which describe
the properties of {\it averaged} forces and flows, and to teach them
about the dynamical fluctuations, away from those average values, for
systems containing small numbers of particles.

In this article, we describe a very simple way to teach these
principles. We start from the ``principle of Maximum Caliber",
first described by E. T. Jaynes.~\citep{jaynes80} It aims to
provide the same type of foundation for dynamics of
many-degree-of-freedom systems that the second law of
thermodynamics provides for equilibria of such systems. To
illustrate the principle, we use a slight variant of one of the
oldest and simplest models in statistical mechanics, the Dog-Flea
Model, or Two-Urn Model.~\citep{klein56, emch02}
Courses in dynamics often introduce Fick's law, Fourier's law, and
the Newtonian-fluid model as phenomenological laws, rather than
deriving them from some deeper foundation. Here, instead, we
describe a simple unified perspective that we have found useful
for teaching these laws from a foundation in statistical dynamics.
In analogy with the role of {\it microstates} as a basis for the
properties of equilibria, we focus on {\it microtrajectories} as
the basis for predicting dynamics.
One argument that might be leveled against the kind of framework
presented here is that in the cases considered here, it is not clear
that it leads to anything different from what one obtains using
conventional nonequilibrium thinking. On the other hand, often,
restating the same physical result in different language can provide a
better starting point for subsequent reasoning. This point was well
articulated by Feynman in his Nobel lecture~\cite{feynman98} who
noted: ``Theories of the known, which are described by different
physical ideas may be equivalent in all their predictions and are
hence scientifically indistinguishable. However, they are not
psychologically identical when trying to move from that base into the
unknown. For different views suggest different kinds of modifications
which might be made and hence are not equivalent in the hypotheses one
generates from them in one's attempt to understand what is not yet
understood."

We begin with the main principle embodied in Fick's Law.  Why do
particles and molecules in solution flow from regions of high
concentration toward regions of low concentration? To keep it
simple, we consider one-dimensional diffusion along a coordinate
$x$. This is described by Fick's first law of particle
transport,~\cite{rief65, dill03} which says that the average flux,
$\langle J \rangle$, is given in terms of the gradient of average
concentration, $ \partial \langle c \rangle /\partial x$, by
\begin{equation}
\langle J \rangle = -D \frac{\partial \langle c \rangle}{\partial x}
\end{equation}
where $D$ is the diffusion coefficient.  In order to clearly
distinguish quantities that are dynamical averages from those that
are not, we indicate the averaged quantities explicitly by
brackets, $\langle \ldots \rangle$. We describe the nature of this
averaging below, and the nature of the dynamical distribution
functions over which the averages are taken.  But first, we
briefly review the standard derivation of the diffusion equation.
Combining Fick's first law with particle conservation,
\begin{equation}
\label{conteq}
\frac{\partial \langle c \rangle}{\partial t} =  -\frac{\partial \langle J \rangle}{\partial x},
\end{equation}
gives Fick's second law, also known as the diffusion equation:
\begin{equation}
\label{diffeq}
\frac{\partial \langle c \rangle}{\partial t} = D \frac{\partial^2
\langle c \rangle}{\partial x^2}.
\end{equation}
Solving Eq.~(\ref{diffeq}) subject to two boundary conditions and
one initial condition gives both $\langle c(x,t) \rangle$, the average
concentration in time and space, and the average flux $ \langle J(x,t)
\rangle $, when no other forces are present. The generalization to
situations involving additional applied forces is the Smoluchowski
equation.~\cite{dill03}

\begin{figure}[h!]
\centering
\vspace*{0.5cm}
\centering
\includegraphics[width = 10 cm]{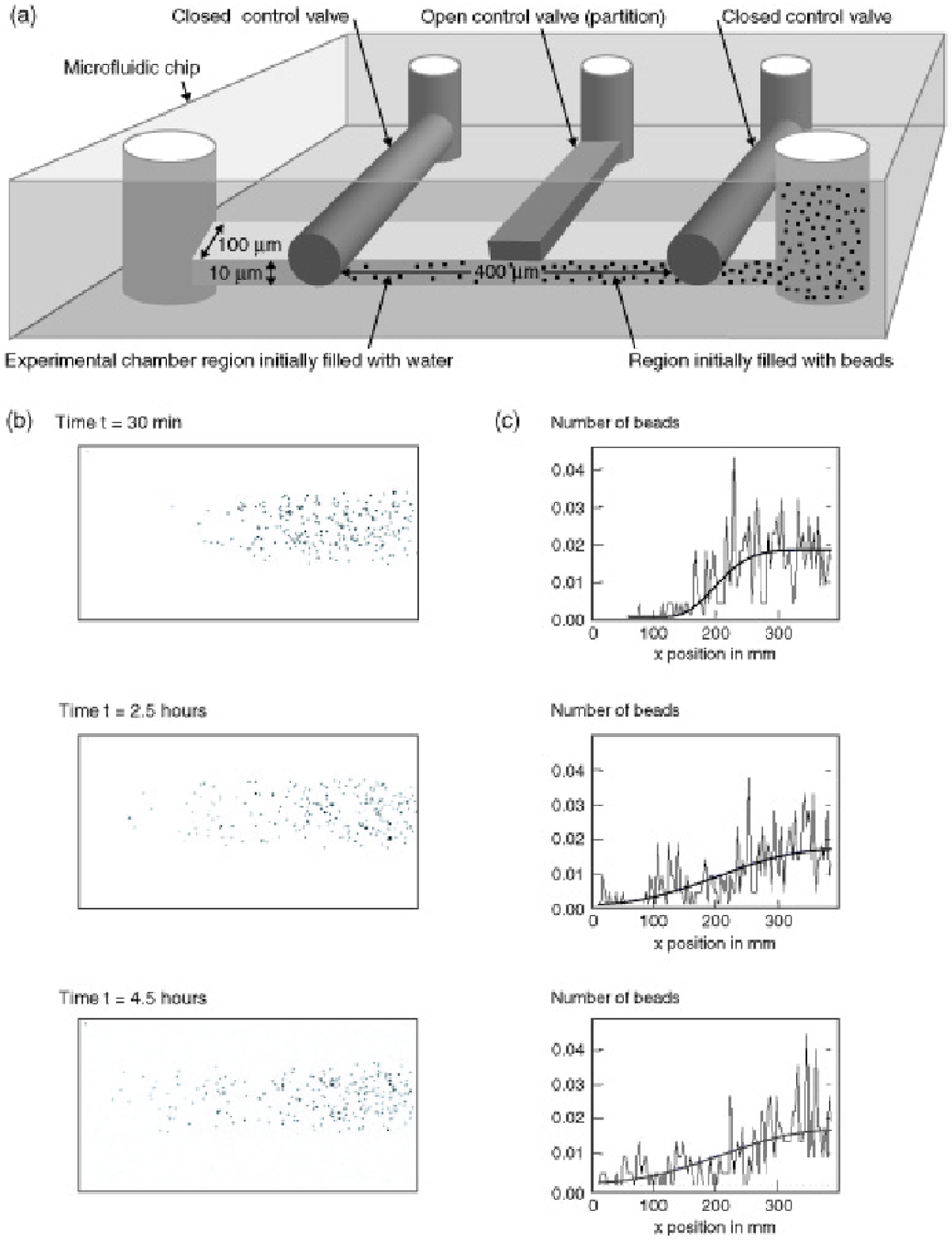}
\caption{\label{FrossoFig}Colloidal free expansion setup to
illustrate diffusion involving small numbers of particles.  (a)
Schematic of experimental setup (see text for details.) (b)
Several snapshots from the experiment. (c) Normalized histogram of
particle positions during the experiment. The solution to the
diffusion equation for the microfluidic ``free expansion"
experiment is superposed for comparison.}
\end{figure}

A simple experiment shows the distinction between {\it averaged}
quantities vs. {\it individual microscopic realizations}. Using a
microfluidics chip like that shown in Fig.~(\ref{FrossoFig}a), it
is possible to create a small fluid chamber divided into two
regions by control valves. The chamber is filled on one side with
a solution containing a small concentration of micron-scale
colloidal particles. The other region contains just water. The
three control valves on top of that microfluidic chamber serve two
purposes: The two outer ones are used for isolation so that no
particles can diffuse in and out of the chamber, while the middle
control valve provides the partition between the two regions. The
time evolution of the system is then monitored after the removal
of the partition (see Fig.~(\ref{FrossoFig}b)). The time-dependent
particle density is determined by dividing the chamber into a
number of equal-sized boxes along the long direction and by
computing histograms of the numbers of particles in each slice as
a function of time. This is a colloidal solution analog of the gas
diffusion experiments of classical thermodynamics. The
corresponding theoretical model usually used is the diffusion
equation. Fig.~({\ref{FrossoFig}c}) shows the solution to the
diffusion equation, as a function of time, for the geometry of the
microfluidics chip. The initial condition is a step function in
concentration at $x = 200\rm{\mu m}$ at time $t=0$.

We use this simple experiment to illustrate one main point.  The
key distinction is that the theoretical curves are very smooth,
while there are very large fluctuations in the experimentally
observed dynamics of the particle densities.  The fluctuations are
large because the number of colloidal particles is small, tens to
hundreds. The experimental data shows that the particle
concentration $c(x, t)$ is a highly fluctuating quantity. It is
not well described by the standard smoothed curves that are
calculated from the diffusion equation.  Of course, when averaged
over many trajectories or when particles are at high
concentrations, the experimental data should approach the smoothed
curves that are predicted by the classical diffusion equation. Our
aim here is to derive Fick's Law and other phenomenological
transport relations at a microscopic level, so that we can
consider both the behaviors of average properties and the
fluctuations, i.e., the dynamical distribution functions, and to
illustrate the Maximum Caliber approach.

\section{The Equilibrium Principle of Maximum Entropy}

We are interested here in dynamics, not statics.  However, our
strategy follows so closely the Jaynes derivation of the Boltzmann
distribution law of equilibrium statistical
mechanics,~\cite{dill03, jaynes57} that we first show the
equilibrium treatment.  To derive the Boltzmann law, we start with
a set of equilibrium microstates $i = 1, 2, 3, \ldots N$ that are
relevant to the problem at hand.  We aim to compute the
probabilities $p_i$ of those microstates in equilibrium.  We
define the entropy, $S$, of the system as
\begin{equation}
S(\{p_i\})=-k_B\sum_{i=1}^N p_i \mbox{ln} p_i,\label{eqprincip1}
\end{equation}
where $k_B$ is the Boltzmann constant. The equilibrium probabilities,
$p_i = p_i^*$ are those values of $p_i$ that cause the entropy to be
maximal, subject to two constraints:
\begin{equation}
\sum_{i=1}^N p_i = 1,\label{eqprincip2}
\end{equation}
which is a normalization condition that insures that the probabilities $p_i$'s sum to one, and
\begin{equation}
\average{E}=\sum_i p_i E_i,\label{eqprincip3}
\end{equation}
which says that the energies, when averaged over all the
microstates, sum to the macroscopically observable average energy.

Using Lagrange multipliers $\lambda$  and $\beta$ to enforce the
first and second constraints, respectively, leads to an expression
for the values $p_i^*$ that maximize the entropy:
\begin{equation}
\sum_{i} [-1 - \mbox{ln} p_i^* -\lambda -\beta E_i ]  = 0.\label{eqprincip4}
\end{equation}
The result is that
\begin{equation}
p_i^*={e^{-\beta E_i} \over Z},
\end{equation}
where $Z$ is the partition function, defined by
\begin{equation}
Z=  \sum_{i} e^{-\beta E_i}.
\end{equation}
After a few thermodynamic arguments, the Lagrange multiplier $\beta$ can be
 shown to be equal to $1/k_{B}T$.~\cite{jaynes57}
This derivation, first given in this simple form by
Jaynes,~\cite{jaynes57} identifies the probabilities that are
both consistent with the observable average energy and that
otherwise maximize the entropy. Jaynes justified this strategy on
the grounds that it would be the best prediction that an observer
could make, given the observable, if the observer is ignorant of all else.
In this case, the observable is the average energy.  While this derivation of the Boltzmann law
is now quite popular, its interpretation as a method of
prediction, rather than as a method of physics, is controversial.  Nevertheless, for our purposes here, it does not
matter whether we regard this as a description of physical systems
or as a strategy for making predictions.

Now, we switch from equilibrium to dynamics, but we use a similar
strategy.  We switch from the Principle of Maximum Entropy to what
Jaynes called the Principle of Maximum Caliber.~\cite{jaynes80} In
particular, rather than focusing on the probability distribution
$p(E_i)$ for the various microstates, we seek
$p[\{\sigma_i(t)\}]$, where $\sigma_i(t)$ is the $i^{th}$
microscopic trajectory of the system.   Again we maximize an
entropy-like quantity, obtained from $p[\{\sigma_{i}(t)\}]$, to
obtain the predicted distribution of microtrajectories.  If there
are no constraints, this maximization results in the prediction
that all the possible microtrajectories are equally likely during
the dynamical process.  In contrast, certain microtrajectories
will be favored if there are dynamical constraints, such as may be
specified in terms of the average flux.

In the following section, we use the Maximum Caliber strategy to
derive Fick's Law using the Dog-Flea model, which is one of the
simplest models that contain the physics of interest.

\section{Fick's Law from the Dog-Flea Model}

\begin{figure}[h!]
\begin{center}
\vspace*{0.5cm}
\includegraphics{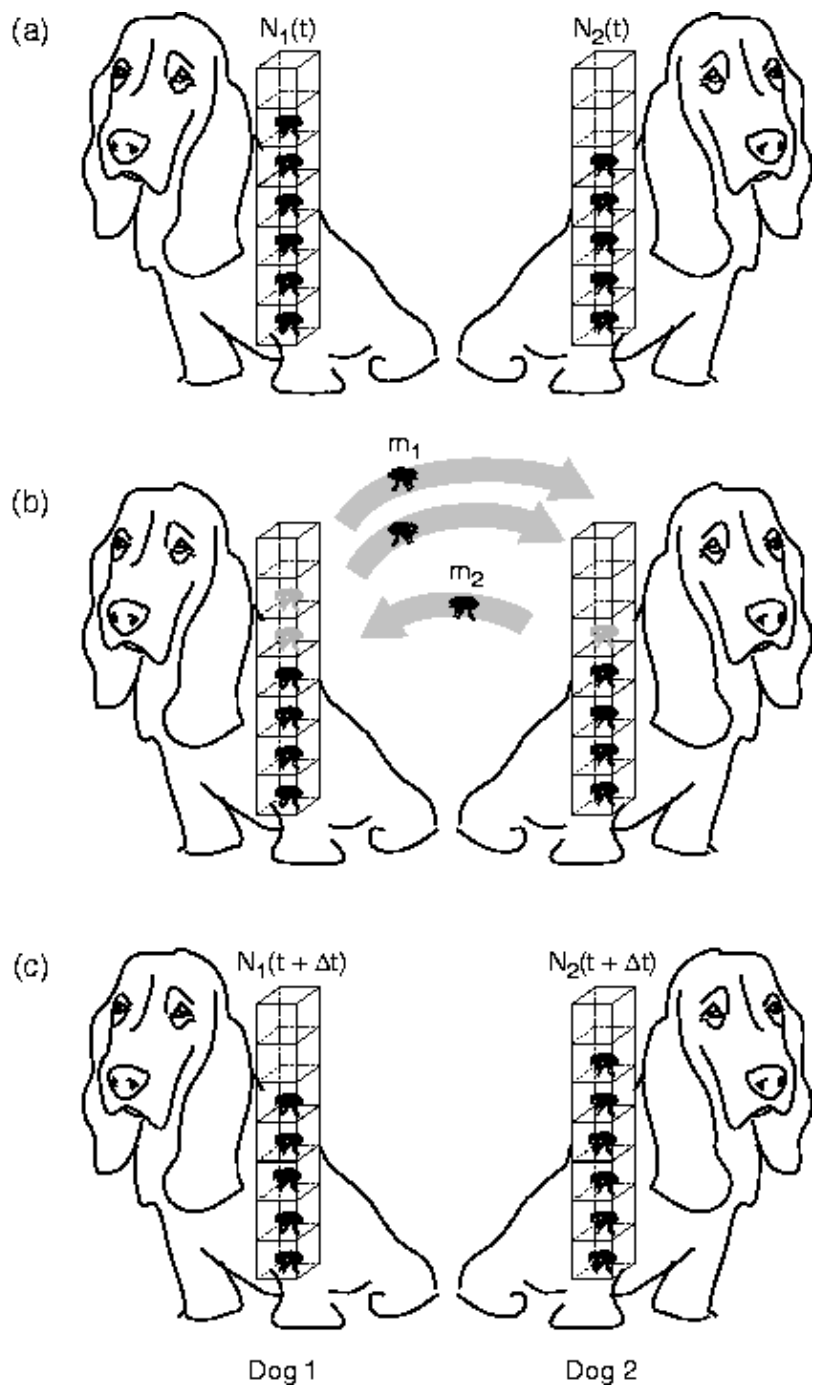}
\caption{\label{DogFleaSchematic}Schematic of the simple dog-flea
  model. (a) State of the system at time $t$, (b) a particular microtrajectory in which two-fleas jump from the dog on the left and one jumps from
  the dog on the right, (c) occupancies of the dogs at time $t+\Delta t$.}
\end{center}
\end{figure}

We want to determine the diffusive time evolution of particles in
a one-dimensional system. The key features of this system are
revealed by considering two columns of particles separated by a
plane, as shown in Fig.~(\ref{DogFleaSchematic}). The left-hand
column (1) has $N_1(t)$ particles at time $t$ and the right-hand
column (2) has $N_2(t)$ particles. This is a simple variant of the
famous ``Dog-Flea" model of the Ehrenfest's introduced in
1907.~\cite{klein56, emch02} Column (1) corresponds to Dog (1), which
has $N_1$ fleas on its back at time $t$, and column (2) corresponds
to Dog (2), which has $N_2$ fleas at time $t$. In any time interval
between time $t$ and $t + \Delta t$, any flea can either stay on
its current dog, or it can jump to the other dog. This model has
been used extensively to study the Boltzmann H-theorem and to
understand how the time asymmetry of diffusion processes arises
from an underlying time symmetry in the laws of
motion.~\citep{kac47,klein56,emch02} Our model is used
for a slightly different purpose.  In particular, our aim is to
take a well-characterized problem like diffusion and to reveal how
the Principle of Maximum Caliber may be used in a concrete way. We
follow the conventional definition of flux, of a number of
particles transferred per unit time, and per unit area. For
simplicity, we take the cross-sectional area to be unity.

First, consider the equilibrium state of our Dog-Flea model.
The total number of ways of partitioning the $(N_1 + N_2)$ fleas (particles) is $W(N_1,
N_2)$,
\begin{equation}
W (N_1, N_2) = \frac{(N_1 + N_2)!}{N_1! N_2!}.
\end{equation}
The state of equilibrium is that for which the entropy, $S = k_B
\ln W$ is maximal. A simple
calculation shows that the entropy is maximal when the value $N_1
= N_1^*$ is as nearly equal to $N_2 = N_2^*$ as possible. In short, at
equilibrium, both dogs will have approximately the same number of
fleas, in the absence of any bias.

Our focus here is on the dynamics -- on how the system reaches that
state of equilibrium.  We discretize time into a series of intervals $\Delta t$. We define a dynamical quantity
$p$, which is the probability that a particle (flea) jumps from one
column (dog) to the other in any time interval $\Delta t$.  Thus, the
probability that a flea stays on its dog during that time interval is
$q=1-p$.  We assume that $p$ is independent of time
$t$, and that all the fleas and jumps are independent of each other.

In equilibrium statistical mechanics, the focus is on the
microstates. However, for dynamics we focus on {\it processes},
which, at the microscopic level, we call the {\it
microtrajectory}. Characterizing the dynamics requires more than
just information about the microstates; we must also consider the
processes. Let $m_1$ represent the number of particles that jump
from column (1) to (2), and $m_2$ is the number of particles that jump
from column (2) to (1), between time $t$ and $t + \Delta t$. There are
many possible different values of $m_1$ and $m_2$: it is possible
that no fleas will jump during the interval $\Delta t$, or that
all the fleas will jump, or that the number of fleas jumping will
be in between those limits. Each one of these different situations
corresponds to a distinct microtrajectory of the system in this
idealized dynamical model. We need a principle to tell us what
number of fleas will jump during the time interval $\Delta t$ at
time $t$. Because the dynamics of this model is so simple, the
implementation of the caliber idea is reduced to nothing more than
a simple exercise in enumeration and counting using the binomial
distribution.

\subsection{The Dynamical Principle of Maximum Caliber}

The probability, $W_d (m_1,m_2|N_1,N_2)$, that $m_1$ particles
jump to the right and that $m_2$ particles jump to the left in a
discrete unit of time $\Delta t$, given that there are $N_1(t)$
and $N_2(t)$ fleas on the dogs at time $t$ is
\begin{equation}
\label{combieq}
\displaystyle{W_d(m_1,m_2|N_1(t),N_2(t)) =\underbrace{\left[p^{m_1}q^{N_1-
m_1}\frac{N_1!}{m_1!(N_1-m_1)!} \right]}_{W_{d_1}} \underbrace{\left[p^{m_2} q^{N_2-
m_2}
\frac{N_2!}{m_2!(N_2-m_2)!} \right]}_{W_{d_2}}} .
\end{equation}
$W_d$ is a count of microtrajectories in dynamics problems in the same vein that
$W$ counts microstates for equilibrium problems.  In the same spirit that the
Second Law of Thermodynamics says to maximize $W$ to predict states of equilibrium,
now for dynamics, we maximize $W_d$ over all the possible
microtrajectories (i.e. over $m_1$ and $m_2$) to predict the
fluxes of fleas between the dogs.
This is the implementation of the Principle of Maximum
Caliber within this simple model.
Maximizing $W_d$ over all the possible processes (different values of
$m_1$ and $m_2$) gives our prediction (right flux $m_1 = m_1^*$ and left flux $m_2 = m_2^*$) for the macroscopic flux that we should observe in experiments.

Since the jumps of the fleas from each dog are independent, we find
our predicted macroscopic dynamics by maximizing $W_{d_1}$ and
$W_{d_2}$ separately, or for convenience their logarithms:
\begin{equation}
\displaystyle{\left. \frac{\partial \ln W_{d_i}}{\partial m_i}\right|_{N,
    m_i=m_{i}^{*}}= 0}, \mbox{\indent}i = 1,2.
\end{equation}
Note that applying Stirling's approximation to Eq.~(\ref{combieq})
$W_d$ gives:
\begin{equation}
\ln W_{d_i} =m_i\ln p + (N_i-m_i)\ln q + N_i\ln N_i  -
m_i\ln m_i  - (N_i - m_i)\ln(N_i-m_i).
\end{equation}
We call $\cal{C}$$=\ln W_d$ the {\it caliber}.  Maximizing $\cal C$ with respect to
$m$ gives
\begin{equation}
\frac{\partial \ln W_{d_i}}{\partial m_i} = \ln p - \ln q  - \ln m_{i}^* +\ln(N_i-m_{i}^*)  = 0.
\end{equation}
This result may be simplified to yield
\begin{equation}
\displaystyle{\ln \left( \frac{m_{i}^*}{N_i - m_{i}^*} \right) = \ln \left(
\frac{p}{1 - p}\right)} ,
\end{equation}
which implies that the most probable jump number is simply given
by
\begin{equation}
m_{i}^* = pN_i.
\end{equation}
But since our probability distribution $W_d$ is nearly symmetric about the most
probable value of flux,  the average number and the most probable
number are approximately the same. Hence, the average net flux to
the right will be,
\[
\langle J(t)\rangle  = {m_1^* - m_2^*\over \Delta t} = p\left[{N_1(t) - N_2(t)\over
    \Delta t}\right] \approx - {p \Delta x^2\over \Delta t}
{\Delta c(x,t) \over \Delta x}
\]
which is Fick's law, in this simple model, and where the diffusion
constant is given by $D=p \Delta x^2/\Delta t$. We have rewritten
$N_1 - N_2 = -\Delta c \Delta x$.

Hence we have a simple explanation for why there is a net flux of
particles diffusing across a plane down a concentration gradient:
more microscopic trajectories lead downhill than uphill.  It shows
that the diffusion constant $D$ is a measure of the jump rate $p$.
This simple model does not make any assumptions that the system is
``near-equilibrium", i.e., utilizing the Boltzmann distribution
law, for example, and thus it indicates that Fick's Law ought also
to apply far from equilibrium.  For example, we could have
imagined that for very steep gradients, Fick's Law might have been
only an approximation and that diffusion is more accurately
represented as a series expansion of higher derivatives of the
gradient.  But, at least within the present model, Fick's Law is a
general result which emerges from counting up microtrajectories.
On the other hand,  we would expect  Fick's law to break down when
the particle density becomes so high that the particles start
interacting with each other thus spoiling the assumption of
independent particle jumps.

\subsection{Fluctuations in Diffusion}

Above, we have shown that the most probable number of fleas that
jump from dog (1) to dog (2) between time $t$ and  $t + \Delta t$ is
$m_1^* = p N_1(t)$.  The model also tells us that sometimes we
will have fewer fleas jumping during that time interval, and
sometimes we will have more fleas.  These variations are a
reflection of the fluctuations resulting from the system following
different microscopic pathways.

We focus now on predicting the fluctuations. To illustrate, let us
first make up a table of $W_d$, the different numbers of possible
microtrajectories, taken over all the values of $m_1$ and $m_2$
(Table~(\ref{tab:tableone})). To keep this illustration simple, let us
consider the following particular case: $N_1(t) = 4$ and $N_2(t) =2$.
Let us also assume $p=q=1/2$. Here, then, are the multiplicities of
all the possible routes of flea flow. A given entry tells how many
microtrajectories correspond to the given choice of $m_1$ and $m_2$.

Notice first that the table confirms our previous discussion.  The
dynamical process for which  $W_d$ is maximal (12
microtrajectories, in this case), occurs when $m_1^* = p N_1 =
1/2 \times  4 = 2$, and $m_2^* = p N_2 = 1/2 \times 2 = 1$.  You can compute
the probability of that particular flux by dividing $W_d = 12$ by
the sum of entries in this table, which is $2^6 = 64$ the total number
of microtrajectories.  The result, which is the fraction of
all the possible microtrajectories that have $m_1^* = 2$ and
$m_2^* = 1$, is $0.18$.  We have chosen an example in which the particle
numbers are very small, so the fluctuations are large; they
account for more than 80 percent of the flow.  In systems having
large numbers of particles, the relative fluctuations are much
smaller than this.

Now look at the top right corner of this table.  This entry says
that there is a probability of 1/64 that both fleas on dog (2)  will
jump to the left while no fleas will jump to the right, implying
that the net flux, for that microtrajectory, is actually
backwards, relative to the concentration gradient.  We call these
``bad actor'' microtrajectories.   In those cases, particles flow
to {\it increase} the concentration gradient, not decrease it.
Traditionally, ``Maxwell's Demon" was an imaginary microscopic
being that was invoked in similar situations in heat flow
processes, i.e., where heat would flow from a colder object to
heat up a hotter one, albeit with low probability.~\citep{gamow88}
In particular, the Demon was supposed to capture the bad actor
microtrajectories.  At time $t$, there are 4 fleas on the left
dog, and 2 on the right. At the next instant in time, $t + \Delta
t$, all 6 fleas are on the left dog, and no fleas are on the
right-hand dog. Notice that this is not a violation of the Second
Law, which is a tendency towards maximum entropy, because the
Second Law is only a statement that the average flow must increase
the entropy; it says nothing about the fluctuations.

Similarly, if you look at the bottom left corner of the table, you
see a regime of ``superflux": a net flux of 4 particles  to the
right, whereas Fick's Law predicts a net flow of only 2 particles
to the right. This table illustrates that Fick's Law is only a
description of the {\it average} or  {\it most probable} flow, and
it shows that Fick's Law is not always exactly correct at the
microscopic level. However, such violations of Fick's Law are of
low probability, a point that we will make more quantitative
below. Such fluctuations have been experimentally measured in
small systems.~\cite{wang02}

We can further elaborate on the fluctuations by defining the
``potencies" of the microtrajectories. We define the potency to be the
fraction of all the trajectories that lead to a substantial change in
the macrostate. The potencies of trajectories depend upon how far the
system is from equilibrium. To see this, let us continue with our
simple system having 6 particles. The total number of microscopic
trajectories available to this system at each instant in our discrete
time picture is $2^6=64$. Suppose that at $t=0$ all 6 of these
particles are in dog (1). The total number of microscopic trajectories
available to the system can be classified once again using $m_1$ and
$m_2$, where in this case $m_2=0$ since there are no fleas on dog (2)
(See Table.~(\ref{tab:tabletwo})).

What fraction of all microtrajectories changes
the occupancies of both dogs by more than some threshold
value, say $\Delta N_i > 1$? In this case, we find that 57 of the 64 microtrajectories cause a change greater than this to the current state.  We call
these potent trajectories.
Now, let us look at the potencies of the same system of $6$ particles
in a different situation, $N_1= N_2= 3$ when the system is in
macroscopic equilibrium (see Table.~(\ref{tab:tablethree})). Here,
only the trajectories with $(m_1, m_2)$ pairs given by $(0,2), (0,3),
(1,3), (2,0), (3,0)$, and $(3,1)$ satisfy our criterion. Summing over
all of these outcomes shows that just $14$ of the $64$ trajectories
are potent in this case.

There are two key observations conveyed by these arguments. First, for
a system far from equilibrium, the vast majority of trajectories at
that time $t$ are potent, and move the system significantly away from
its current macrostate. Second, when the system is near equilibrium,
the vast majority of microtrajectories leave the macrostate unchanged.
Let us now generalize from the tables above, to see when fluctuations
will be important.

\subsubsection{Fluctuations and Potencies}

A simple way to characterize the magnitude of the fluctuations is to look at the
width of the $W_d$ distribution.~\cite{dill03}  It is shown in standard
statistics texts that for a binomial distribution such as ours,
for which the mean and most probable value both equal $ m_{i}^* = Np_{i}$,
the variance is $\sigma_{i}^2 = N_{i}pq$.  The variance characterizes the
width.  Moreover, if $N_{i}$ is sufficiently large, a binomial
distribution can be well-approximated by a Gaussian distribution
\begin{equation}
\displaystyle{{\cal P}(m_i, N_i) = \frac{1}{\sqrt{2 \pi N_ipq}}
\exp{\left(-\frac{(m_i - N_ip)^2}{2 N_ipq} \right)} },
\end{equation}
an approximation we find convenient since it leads to simple
analytic results.   However, this distribution function is not
quite the one we want. We are interested in the distribution of
flux, $P(J) = P(m_1 - m_2)$, not the distribution of right-jumps
$m_1$ or left-jumps $m_2$ alone, ${\cal P}(m)$.

However, due to a remarkable property of the Gaussian
distribution, it is simple to compute the quantity we want.  If
you have two Gaussian distributions, one with mean $\langle x_1
\rangle$ and variance $\sigma_1^2$, and the other with mean
$\langle x_2 \rangle$ and variance $\sigma_2^2$, then the
distribution function, $P(x_1 - x_2)$ for the difference will also
be a Gaussian distribution, having mean $\langle x_1\rangle -
\langle x_2 \rangle$ and variance $\sigma^2 = \sigma_1^2 +
\sigma_2^2$.

For our binomial distributions, the means are $m_1^* = pN_1$ and $m_2^* = pN_2$ and
the variances are $\sigma_1^2 = N_1 pq$ and $\sigma_2^2 = N_2 pq$, so the distribution
of the net flux, $J = m_1 - m_2$ is
\begin{equation}
\displaystyle{ P(J) = \frac{1}{\sqrt{2\pi (pqN)}} \exp\left(-
\frac{(J - p(N_1-N_2))^2}{2pqN}\right) } \label{FluxProbability1},
\end{equation}
where $N = N_1 + N_2$.

\begin{figure}[h!]
\vspace*{0.5cm}
\includegraphics{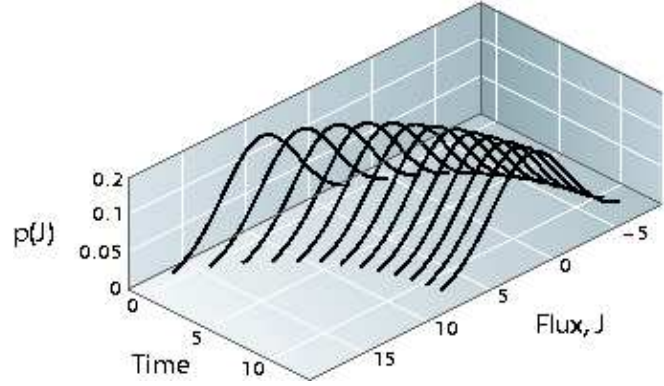}
\caption{\label{FluxDist}Schematic of the distribution of fluxes for different
time points as the system approaches equilibrium.}
\end{figure}

Figure~(\ref{FluxDist}) shows an example of the distributions of fluxes at
different times, using $p = 0.1$, and starting from $N_1 = 100, N_2 =
0$. We update each time step using an averaging scheme, $N_1(t +
\Delta t) = N_1(t) - N_1(t) p + N_2(t) p$. The figure shows how the
mean flux is large at first and decays towards equilibrium, $J = 0$.
This result could also have been predicted from the diffusion
equation. However, equally interesting are the wings of the
distributions, which show the deviations from the average flux, and
these are not predictable from the diffusion equation. One measure of
the importance of the fluctuations in a given dynamical problem is the
ratio of the standard deviation $\sigma$, to the mean,
\begin{equation}
{\sqrt{\sigma^2}\over {J}} = {\sqrt{Npq}\over{(N_1 - N_2)p}} .
\end{equation}
In the limit of large $N_1$, the above reduces to,
\begin{equation}
{\sqrt{\sigma^2}\over {J}} = {\sqrt{Npq}\over{(N_1 - N_2)p}} \sim N^{-1/2} .
\end{equation}

In a typical bulk experiment, particle numbers are large, of the order
of Avogadro's number $10^{23}$. In such cases, the width of the flux
distribution is exceedingly small and it becomes overwhelmingly
probable that the mean flux will be governed by Fick's law. However,
within biological cells and in applications involving small numbers of
particles, the variance of the flux can become significant. It has
been observed that both rotary and translational single motor proteins
sometimes transiently step backwards, relative to their main direction
of motion.~\cite{howard01}

As a measure of the fluctuations, we now calculate the variance in the flux.  It follows from
Eq.~(\ref{FluxProbability1}) that
$\average{J^2} =  Npq$, where $N = N_1 + N_2$.  Thus, we can represent the magnitude of the fluctuations as $\delta$,
\[\delta = \sqrt{\frac{\langle (\Delta J)^2 \rangle}{\langle J \rangle^2}} =
\frac{\sqrt{Npq}}{pfN} \propto \frac{1}{f} \sqrt{\frac{q}{p} N^{-1}},
\]
where the quantity $N=N_1+N_2$ is the total number of fleas and
$f=(N_1-N_2)/(N)$, the normalized concentration difference.  The quantity $\delta$ is also a measure of the degree of backflux. In
the limit of large $N$, $\delta$ goes to zero.  That is, the noise diminishes with
system size. However, even when $N$ is large, $\delta$ can still be large (indicating the
possibility of backflux) if the concentration gradient, $N_1-N_2$, is  small.

\begin{figure}[h!]
\begin{center}
\vspace*{0.5cm}
\includegraphics{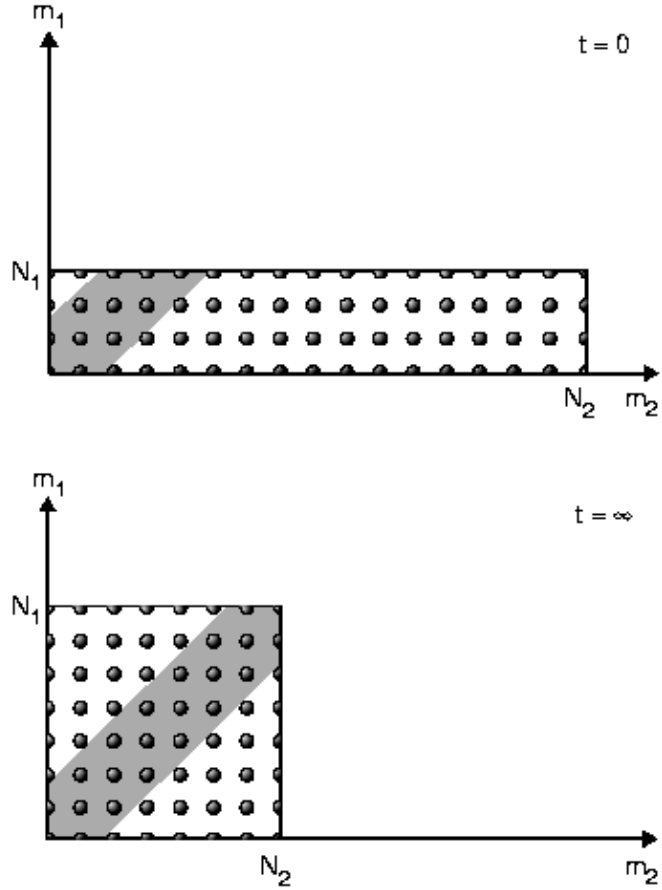}
\caption{\label{fig:potencyillust}Schematic of which trajectories are
potent and which are impotent.  The shaded region corresponds to
the impotent trajectories for which $m_1$ and $m_2$ are either equal or
approximately equal and hence make relatively small change in the
macrostate. The unshaded region corresponds to potent trajectories.}
 \end{center}
\end{figure}

\begin{figure}[h!]
\begin{center}
\vspace*{0.5cm}
\includegraphics{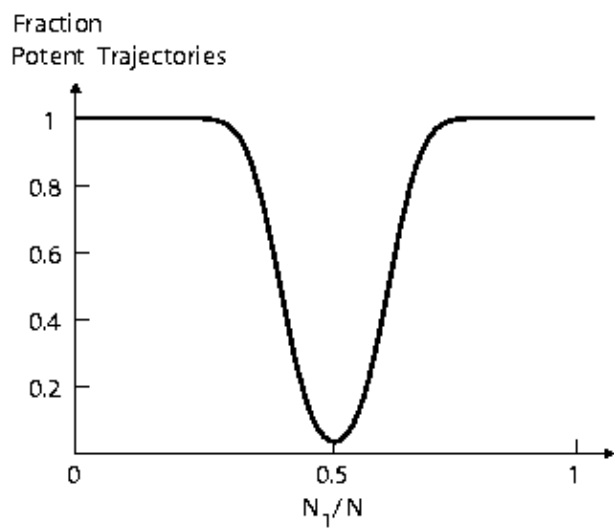}
\caption{\label{fig:potency}Illustration of the potency of the
  microtrajectories associated with different distributions of $N$ particles on the
two dogs. The total number of particles $N_1 + N_2 = N = 100$. }
 \end{center}
\end{figure}

Let us look now at our other measure of fluctuations, the potency.
Trajectories that are {\it not} potent should have $|m_1-m_2|
\approx 0$ which corresponds to a negligible change in the current
state of the system as a result of a given microtrajectory. In
Fig.~(\ref{fig:potencyillust}), the impotent microtrajectories are
shown as the shaded band for which $m_1 \approx m_2$. To quantify
this, we define impotent trajectories as those for which
$|m_1-m_2| \le h$, ($h \ll N$). In the Gaussian model, the
fraction of impotent trajectories is
\bea \impotent &\approx&
\int_{-h}^{h}dJ {1\over \sqrt{2\pi
      Npq}}\exp\left({-(J - (N_1-N_2)p)^{2}\over 2 Npq}\right)\\ & = &
      {1\over2}\left(\textrm{erf}\left[{h+(N_1-N_2)p\over
      \sqrt{2Npq}}\right] + \textrm{erf}\left[{h-(N_1-N_2)p\over
      \sqrt{2Npq}}\right]\right),
      \label{eq:impotentappr}
\eea
and corresponds to summing over the subset of trajectories that have a small
flux.  To keep it simple, we did a computation, taking $p = q = 1/2$, and for which
the expression for the probability distribution for the
microscopic flux $m_1 - m_2$ is given by
Eq.~(\ref{FluxProbability1}).
The choice of $h$ is arbitrary, so let us just choose $h$ to be one
standard deviation, $\sqrt{N/4}$. Fig.~(\ref{fig:potency}) shows
potencies for various values of $N_1$ and $N_2$.  When the
concentration gradient is large, most trajectories are potent,
leading to a statistically significant change of the macrostate,
whereas when the concentration gradient is small, most
trajectories have little effect on the macrostate.

\begin{figure}[h!]
\begin{center}
\vspace*{0.5cm}
\includegraphics{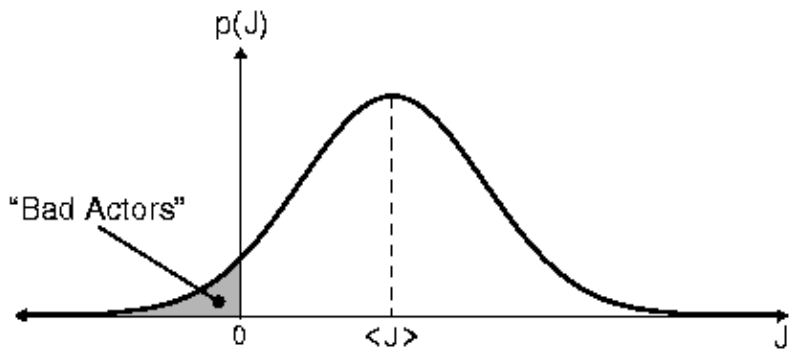}
\caption{\label{fig:tail}Illustration of the notion of bad actors.
Bad actors are essentially the microtrajectories which contribute
net particle motion which has the opposite sign from the
macroflux. }
 \end{center}
\end{figure}

\begin{figure}[h!]
\begin{center}
\vspace*{0.5cm}
\includegraphics{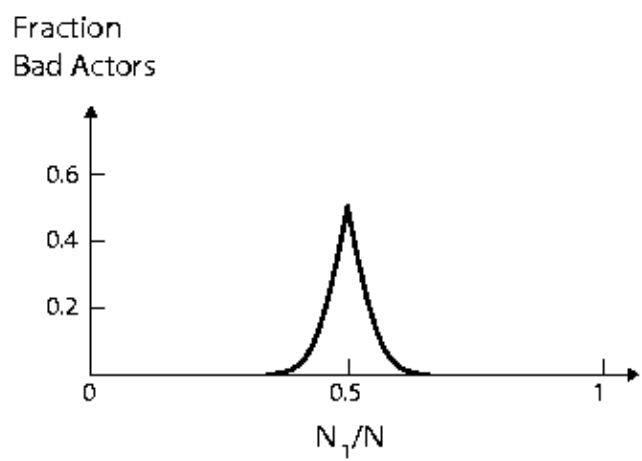}
\caption{\label{BadActors}The fraction of all possible trajectories that go
  against the direction of the macroflux, for $N = 100$. The fraction of bad actors is highest at $N_1 = N/2 = 50$.}
 \end{center}
\end{figure}

As another measure of fluctuations, let us now consider the ``bad actors'' (see Fig.~(\ref{fig:tail})).
If the average flux is in the direction
from dog (1) to dog (2), what is the probability you will observe flux in the opposite
direction (bad actors)? Using
Eq.~(\ref{FluxProbability1}) for $P(J)$, we get
\bea
\badactors &\approx& \int_{-\infty}^{0} {1\over \sqrt{2\pi
    Npq}}\exp\left({-(J - (N_1-N_2)p)^{2}\over 2 Npq}\right)\\ &=&
       {1\over 2}\left(1- \textrm{erf}\left[{(N_1-N_2)p\over
           \sqrt{2Npq}}\right]\right), N_2 > N_1
\eea
which amounts to summing up the fraction of trajectories for which
$J \le 0$.   Figure~(\ref{BadActors}) shows the fraction of bad actors
for $p = q = 1/2$. Bad actors are rare
when the concentration gradient is large, and highest when the
gradient is small.  The discontinuity in the slope of the curve in
Fig.~(\ref{BadActors}) at $N_1/N = 0.5$ is a reflection of the fact
that the mean flux abruptly changes sign at that value. 

\section{Fourier's Law of Heat Flow}
While {\it particle flow} is driven by concentration gradients,
according to Fick's law, $\langle J \rangle = -D \frac{\partial
c}{\partial x}$, {\it energy flow} is driven by temperature gradients,
according to Fourier's law~\cite{rief65}:
\[
\langle J_q \rangle = -\kappa \frac{\partial  T }{\partial x}.
\]
Here, $J_q$ is the energy
transferred per unit time and per unit cross-sectional area by heat flow and $\partial T/
\partial x$ is the temperature gradient that drives it, indicated here for the one-dimensional case.  $\kappa$, the thermal
conductivity,~\cite{rief65} plays the role that the diffusion
coefficient plays in Fick's Law.

To explore Fourier's law, we return to the Dog-Flea model as described
in Sec.~III. Now, columns (1) and (2) can differ not only in their
particle numbers, $N_1(t)$ and $N_2(t)$, but also in their
temperatures, $T_1(t)$ and $T_2(t)$. To keep it simple here, we assume
that each column is at thermal equilibrium and that each particle that
jumps carries with it the average energy, $\langle mv^2/2 \rangle =
k_BT/2$ from the column it left. Within this simple model, all energy
is transported by hot or cold molecules switching dogs. Although in
general, heat can also flow by other mechanisms mediated by
collisions, for example, our aim here is just the simplest
illustration of principle. The average heat flow at time $t$ is
\bea
\average{J_q} =  {m_1^*\over \Delta t} (k_BT_1/2)  -
{m_2^*\over \Delta t} (k_BT_2/2) = {pk_B\over \Delta t}[N_1T_1 - N_2T_2]
\eea
where  $m_1^*$ and $m_2^*$ are, as defined in Sec.~IIIA the numbers of
particles jumping from each column at time $t$. If the particle
numbers are identical, $N/2=N_1=N_2$, then
\[
\langle J_q \rangle = {pk_BN \over \Delta t} (T_1 - T_2) = -\kappa {\Delta T\over \Delta x},
\]
which is Fourier's Law for the average heat flux, within this
two-column model.  The model predicts that the thermal
conductivity is $\kappa = (pk_BN \Delta x)/(\Delta t)$, which can be
expressed in a more canonical form as $\kappa = pk_Bnv_{\rm av} \Delta
x$, when written in terms of the particle density $n = N/\Delta x$
and the average velocity, $v_{\rm av} = \Delta x/\Delta t$.  Our
simple model gives the same thermal conductivity as given by the
kinetic theory of gases,~\cite{rief65} $\kappa = (1/2)k_B n v_{\rm av}
l$, where $l$ is the mean free path, if $p\Delta x$ in our model
corresponds to $l/2$, half the mean-free path. Hence, this
simple model captures the main physical features of heat flow,
again by appealing to the idea of summing over the weighted
microtrajectories available to the system.

\section{Newtonian Viscosity}

\begin{figure}[h!]
\begin{center}
\vspace*{0.5cm}
\includegraphics{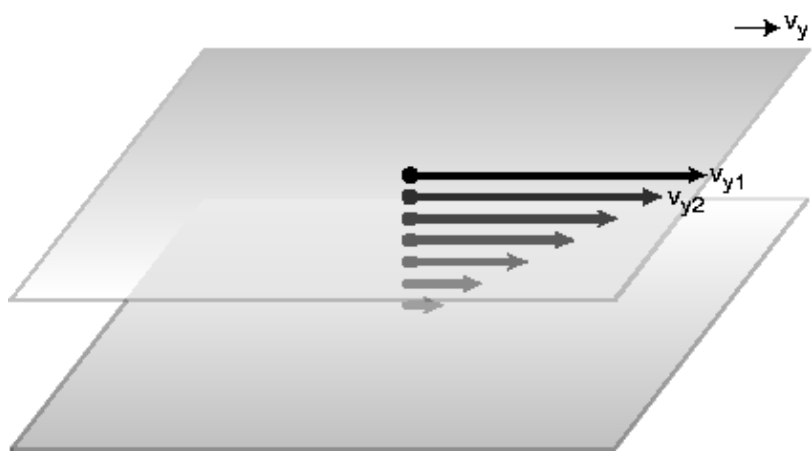}
\caption{\label{fig:newton}Illustration of Newton's law of viscosity.
The fluid is sheared with a constant stress. The fluid velocity
decreases continuously from its maximum value at the top of the fluid
to zero at the bottom. There is thus a gradient in the velocity which
can be related to the shear stress in the fluid.}
 \end{center}
\end{figure}

Another phenomenological law of gradient-driven transport is that of Newtonian
viscosities,~\cite{rief65}
\[
\tau = \eta {dv_y \over dx},
\]
where $\tau$ is the shear stress that is applied to a fluid, $dv_y/dx$ is
the resultant shear rate, and the proportionality coefficient,
$\eta$, is the viscosity of a Newtonian fluid.  Whereas Fick's law
describes particle transport and Fourier's law describes energy
transport, this Newtonian law describes the transport (in the x-direction, from the top moving plate toward the bottom fixed plate)
of linear momentum that acts in the y-direction (parallel to the plates) (see Fig.~(\ref{fig:newton})).
Returning to the Dog-Flea model of Sec.~III, suppose each particle in column
(1) carries momentum $mv_{y_{1}}$ along the y-axis, and that
$m_1^*$ particles hop from column (1) to (2) at time $t$, carrying with them
some linear momentum. As before, we consider
the simplest model that every particle carries the same average momentum
from the column it leaves to its destination column.

The flux, $J_p$, is the amount
of y-axis momentum that is transported from one plane to the next
in the x-direction, per unit
area:
\[
\displaystyle{\langle J_{p} \rangle = {m_1^*\over \Delta t} \left(mv_{y_{1}}\right) -
     {m_2^*\over \Delta t}\left(mv_{y_{2}}\right) = {pm\over \Delta t}\left[ N_1v_{y_{1}} -
N_2v_{y_{2}}\right]}.
\]
If the number of particles is the same in both columns,  $N/2=N_1=N_2$, this simplifies to
\[
\displaystyle{\langle J_{p} \rangle = {pmN\over \Delta
t}\left[v_{y_{1}} - v_{y_{2}}\right] = \eta {\Delta v_y \over \Delta
x}},
\]
which is the Newtonian law of viscosity in this two-column model.
The viscosity is predicted by this model to be $\eta = (pmN \Delta
x)/(\Delta t)$. Converting this to a more canonical form gives
$\eta = p m n v_{av}\Delta x$, where $n = N/\Delta x$ is the
particle density, and $v_{\rm av} = \Delta x/\Delta t$ is the
average velocity. This is equivalent to the value given by the
kinetic theory of gases,~\cite{rief65} $\eta = (1/3) m n l
v_{av}$, if $p\Delta x$ from our model equals $(1/3) l$, one-third
of the mean-free path length. Note that this simple model  based
upon molecular motions will clearly not be applicable to complex
fluids where the underlying molecules possess internal structure.

\section{Chemical Kinetics Within the Dog-Flea Model}

Let us now look at chemical reactions using the Dog-Flea model.
Chemical kinetics can be modeled using the Dog-Flea model when the
fleas have preference for one dog over the other. Consider the reaction
\[
A \underset{k_{r}}{\overset{k_{f}}{\rightleftharpoons}} B.
\]
The time-dependent average concentrations, $[A](t)$ and $[B](t)$
are often described by chemical rate equations,~\cite{dill03}

\begin{eqnarray}
\frac{d[A]}{dt} &=& -k_f [A] + k_r [B]\nonumber \\
\frac{d[B]}{dt} &=&  k_f [A] - k_r [B]
\end{eqnarray}
where $k_f$ is the average conversion rate of an A to a B, and $k_r$ is the
average conversion rate of a B to an A.  These rate expressions describe only
average rates; they do not tell us the distribution of rates.  Some A's will
convert to B's faster than the average rate $k_f [A]$ predicts, and some will
convert more slowly.  Again, we use the Dog-Flea model as a microscopic model
for this process.  We use it to consider both the average concentrations and the fluctuations in concentrations.

Now, dog (1) represents chemical species A and dog (2) represents
chemical species B. The net chemical flux from 1 to 2 is given by
$J_c = m_1^* - m_2^*$.  What is different about our model for
these chemical processes than in our previous situations is that
now the intrinsic jump rate from column 1 (species A), $p_1$, is
different than the jump rate from column 2, $p_2$. This simply
reflects the fact that a forward rate can differ from a backward
rate in a chemical reaction.  Now, fleas have a different escape
rate from each dog.  Fleas escape from Dog (1) at rate $p_1$ and
fleas escape from Dog (2) at rate $p_2$.  Maximizing $W_d$ gives
$m_1^* = N_1 p_1$ and $m_2^* = N_2 p_2$, so the average flux (which is
the almost the same as the most probable flux because of the approximately
symmetric nature of the binomial distribution) at time $t$ is,
\[
\la J \ra = N_1p_1 - N_2p_2 = k_f [A] - k_r [B],
\]
which is just the standard mass-action
rate law, expressed in terms of the mean concentrations.
The mean values satisfy detailed balance at equilibrium ( $\la J \ra$=0 $\Rightarrow N_2/N_1$
= $p_1/p_2$ = $k_{f}/k_{r}$).

\begin{figure}[h!]
\begin{center}
\vspace*{0.5cm}
\includegraphics{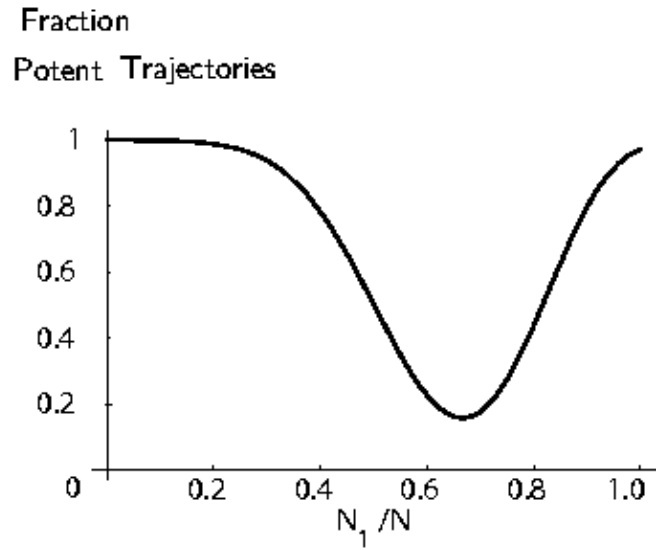}
\caption{\label{fig:chempot}The fraction of potent trajectories,
  $\potent$, as a function of $N_1/N$ for $N_1 + N_2 = N = 100$, when
  $p_1 = 0.1$ and $p_2= 0.2$ are not equal. The minimum value of
  potency does not occur at $N_1/N = 0.5$, but at $N_1/N =
  0.66$. This value of $N_1/N$ also corresponds to its equilibrium value
  given by $p_2/(p_1 + p_2)$.}
 \end{center}
\end{figure}

More interesting than the well-known behavior of the mean chemical
reaction rate is the fluctuational behavior.  For example,
if the number of particles is small,
then even when $k_f [A] - k_r [B] > 0$, indicating an average
conversion of A's to B's, the reverse can happen occasionally
instead. When will these fluctuations be large?
As in Sec.~III, we first determine the probability distribution of
the flux $J$.  In this case, the probability
distribution becomes:
\begin{equation}
\displaystyle{ P(J) = \frac{1}{\sqrt{2\pi (p_1q_1N_1+p_2q_2N_2)}}
\exp\left(- \frac{(J -
(N_1p_1-N_2p_2))^2}{2(p_1q_1N_1+p_2q_2N_2)}\right) },
\label{FluxProbability2}
\end{equation}
Again, let us use this flux distribution function to consider the fluctuations
in the chemical reaction.  The relative variance in the flux is
\[
\frac{\langle (\Delta J)^2\rangle}{(\langle J \rangle)^2} = \frac{\sqrt{N_1p_1q_1 + N_2p_2q_2}}{N_1p_1 - N_2p_2}.
\]
As before, the main message is that when the system is not yet at
equilibrium (i.e., the denominator is non-zero), macroscopically large systems
will have negligibly small fluctuations.  The relative magnitude of fluctuations
scales approximately as $N^{-(1/2)}$.  Let us also look at the potencies of
microtrajectories as another window into fluctuations.  Using Eq.~(\ref{eq:impotentappr}) with $p_1$ and $p_2$ gives the fraction of trajectories
that are impotent as
\bea
\impotent &\approx& \int_{-h}^{h}dJ {1\over \sqrt{2\pi(N_1p_1q_1 + N_2 p_2q_2)}}\exp\left({-(J - (N_1p_1-N_2p_2))^{2}\over 2
(N_1p_1q_1+N_2p_2q_2)}\right)\\ & = &
      {1\over2}\left(\textrm{erf}\left[{h+(N_1p_1-N_2p_2)\over
      \sqrt{2(N_1p_1q_1+N_2p_2q_2)}}\right] + \textrm{erf}\left[{h-(N_1p_1-
N_2p_2)\over
      \sqrt{2(N_1p_1q_1+N_2p_2q_2)}}\right]\right),\label{eq:cheimpappr}
\eea Using $N_1 + N_2  = N = 100$, and $p_1 = 0.1$ and $p_2 =
0.2$, $\potent = 1 - \impotent$ as a function of $N_1/N$ is shown
in Fig.~\ref{fig:chempot}.

\section{Deriving the Dynamical Distribution Function from Maximum Caliber}

Throughout this paper, we have used the binomial distribution
function, $W_d$, as the basis for our treatment of stochastic
dynamics.  The Maximum Caliber idea says that if we find the value of
$W_d$ that is maximal with respect to the microscopic trajectories,
this will give the macroscopically observable flux.  Here, we now
restate this in a more general way, and in terms of the probabilities
of the trajectories.

Let $P(i)$ be the probability of a microtrajectory $i$ during the
interval from time $t$ to $t + \Delta t$. A microtrajectory is a
specific set of fleas that jump; for example microtrajectory $i=27$
might be the situation in which fleas number 4, 8, and 23 jump from
dog (1) to (2).  We take as a constraint the average number of jumps,
$\langle m \rangle$, the macroscopic observable.  The quantity $m_i =
3$ in this case indicates that trajectory $i$ involves 3 fleas
jumping.  We express the caliber $\cal C$ as
\bea
{\cal C} = \sum_i P(i)\ln P(i) - \lambda \sum_i m_i P(i) -\alpha
\sum_i P(i) \label{caliberoneeq}
\eea
where, $\lambda$ is the Lagrange multiplier that enforces the constraint
of the average flux and $\alpha$ is the Lagrange multiplier that enforces the
normalization condition that the $P(i)$'s sum to one.  Maximizing the caliber
gives the populations of the microtrajectories,
\bea
P(i) = \exp (-\alpha -\lambda m_i).
\eea
Note that the probability $P(i)$ of the $i^{\rm th}$ trajectory
depends only on the total number $m_i$ of the jumping fleas. Also, all
trajectories with the same $m_i$ have same probabilities. Now, in the
same way that it is sometimes useful in equilibrium statistical
mechanics to switch from microstates to energy levels, we now express
the population $P(i)$ of a given microtrajectory, instead, in terms of
$\rho(m)$ the fraction of {\it all} the microtrajectories that involve
$m$ jumps during this time interval,
\bea
\rho(m) = g(m) Q(m),
\eea
where, $g(m) = N!/[m! (N-m)!]$ is the ``density of
trajectories'' with flux $m$ (in analogy with the density of
states for equilibrium systems), and $Q(m)$ is the probability $P(i)$ of the
microtrajectory $i$ with $m_i = m$. In other words, $i$ denotes
a microtrajectory (specific set of fleas jumping) while $m$ denotes a
microprocess (the number of fleas jumping).  The total number of $i$'s
associated with a given $m$ is precisely $g(m)$. It can also be easily
seen that,
\bea
\sum_{i}P(i) &=& \sum_{m = 0}^{N}g(m)Q(m) = \sum_{m=0}^{N}\rho(m) = 1\\
\langle m \rangle &=& \sum_i m_i P(i) = \sum_{m=0}^{N}mg(m)Q(m)= \sum_{m=0}^N m \rho(m).
\eea
Thus, the distribution of jump-processes written in terms of the
jump number, $m$, is
\bea
\rho(m) = \frac{N!}{m! (N-m)!} \exp(-\alpha) \exp(-\lambda m).
\label{calibertwoeq}
\eea
The Lagrange multiplier $\alpha$ can be eliminated by summing over
all trajectories and requiring that $\sum_{m_i=0}^N\rho(m_i)= 1$, i.e.,
\bea \exp(\alpha) = \sum_m^N g(m) e^{-\lambda m} = \sum_m^N
\frac{N!}{m!(N-m)!} e^{-\lambda m} = (1 + e^{-\lambda})^N.
\label{Normalization} \eea
Combining Eqs.~(\ref{calibertwoeq}) and (\ref{Normalization}) gives
\begin{equation}
\rho(m) = \frac{N!}{m! (N-m)!} \frac{\exp(-\lambda m)}{(1 +
\exp(-\lambda))^N}.
\label{rhom}
\end{equation}
If we now let
\bea
p= \frac{\exp(-\lambda)}{1 + \exp(-\lambda)} ,
\label{pdef}
\eea
then we get
\bea
p^m = \frac{\exp(-\lambda m)}{(1 + \exp(-\lambda))^m} ,
\label{pdefm}
\eea
and
\bea
(1-p)^{N-m} = \frac{1}{(1 + \exp(-\lambda))^{N-m}} .
\label{qdefm}
\eea
Combining Eqs.~(\ref{rhom}),~(\ref{pdefm}), and (\ref{qdefm}) gives the simple form
\[
\rho(m) = \frac{N!}{m! (N-m)!} p^m(1-p)^{(N-m)},
\]
that we have used throughout this paper in
Eq.~(\ref{combieq}).

\section{Summary and Comments}

We have shown how to derive the phenomenological laws of
nonequilibrium transport, including Fick's law of diffusion, the
Fourier law of heat conduction, the Newtonian law of viscosity, and
mass-action laws of chemical kinetics, from a simple physical
foundation that can be readily taught in elementary courses. We use
the Dog-Flea model, originated by the Ehrenfests, for describing
how particles, energy, or momentum can be transported across a plane.
We combine that model with the Principle of Maximum Caliber, a
dynamical analog of the way the Principle of Maximum Entropy is used
to derive the laws of equilibrium. In particular, according to Maximum
Entropy, you maximize the entropy $S(p_1, p_2, \ldots, p_N)$ with
respect to the probabilities $p_i$ of $N$ microstates, subject to
constraints, such as the requirement that the average energy is known.
That gives the Boltzmann distribution law. Here, for dynamics, we
focus on microtrajectories, rather than microstates, and we maximize a
dynamical entropy-like quantity, subject to an average flux
constraint. In this way, maximizing the caliber is the dynamical
equivalent of minimizing a free energy for predicting equilibria. A
particular value of this approach is that it also gives us fluctuation
information, not just averages. In diffusion, for example, sometimes
the flux can be a little higher or lower than the average value
expected from Fick's Law. These fluctuations can be particularly
important for biology and nanotechnology, where the numbers of
particles can be very small, and therefore where there can be
significant fluctuations in rates, around the average.

\acknowledgements 
It is a pleasure to acknowledge the helpful comments
and discussions with Dave Drabold, Mike Geller, Jan\'{e} Kondev,
Stefan M\"{u}ller, Hong Qian, Darren Segall, Pierre Sens, Jim Sethna,
Ron Siegel, Andrew Spakowitz, Zhen-Gang Wang, and Paul Wiggins. We
would also like to thank Sarina Bromberg for the help with the
figures. KAD and MMI would like to acknowledge support from NIH grant
number R01 GM034993 . RP acknowledges support from NSF grant number
CMS-0301657, the Keck foundation, NSF NIRT grant number CMS-0404031,
and NIH Director's Pioneer Award grant number DP1 OD000217 .

\bibliography{Caliber}

\pagebreak 

\clearpage

\begin{table}
\begin{center}
\begin{tabular}{|l|ccc|}\hline
$\,_{m_1}{\diagdown}^{m_2}$ & 0 & 1 & 2 \\ \hline
0 & 1 & 2 & 1 \\
1 & 4 & 8 & 4 \\
2 & 6 & 12 & 6 \\
3 & 4 & 8 & 4 \\
4 & 1 & 2 & 1 \\ \hline
\end{tabular}
\end{center}
\caption{\label{tab:tableone} Trajectory multiplicity table for the specific case
  where $N_1(t) = 4$ and $N_2(t) = 2$. Each entry in the table
  corresponds to the total number of trajectories for the particular
  values of $m_1$ and $m_2$.}
\end{table}

\begin{table}
\begin{center}
\begin{tabular}{|l|c|}\hline
$\,_{m_1}{\diagdown}^{m_2}$ & 0  \\ \hline
0 & 1  \\
1 & 6 \\
2 & 15 \\
3 & 20\\
4 & 15 \\
5 & 6 \\
6 & 1 \\ \hline
\end{tabular}
\end{center}
\caption{\label{tab:tabletwo} Trajectory multiplicity table for the specific case
  $N_1(t) = 6$ and $N_2(t) = 0$ when the system is far from
  macroscopic equilibrium.}
\end{table}

\begin{table}
\begin{center}
\begin{tabular}{|l|cccc|}\hline
$\,_{m_1}{\diagdown}^{m_2}$ & 0 & 1 & 2 & 3 \\ \hline
0 & 1 & 3 & 3 & 1  \\
1 & 3 & 9 & 9 & 3 \\
2 & 3 & 9 & 9 & 3 \\
3 & 1 & 3 & 3 & 1\\\hline
\end{tabular}
\end{center}
\caption{\label{tab:tablethree} Trajectory multiplicity table for the specific case
  $N_1(t) = 3$ and $N_2(t) = 3$ when the system is at the macroscopic
  equilibrium.}
\end{table}

\end{document}